\documentclass[runningheads]{llncs}

\usepackage{pgfplots}
\pgfplotsset{compat=1.8}
\pgfplotsset{yticklabel style={text width=3em,align=right}}
\pgfplotsset{
    exclusively/.style 2 args={
    x filter/.code={
        \edef\weekofmyrow{\thisrow{#1}}
        \edef\weektocheck{#2}
        \ifx\weekofmyrow\weektocheck
        \else\def\pgfmathresult{nan}
        \fi
    }
    }
}
\newcommand{\oddweekplot}[3]{
\addplot[bar shift=0pt,fill=blue!55!white]
    table[x=ID, y=#2, col sep=comma,
    exclusively={week}{#3}
    ]{#1Result.csv};
}
\newcommand{\evenweekplot}[3]{
\addplot[bar shift=0pt,fill=blue!30!white]
    table[x=ID, y=#2, col sep=comma,
    exclusively={week}{#3}
    ]{#1Result.csv};
}
\newcommand{\addtikz}[2]{
\begin{tikzpicture}
\begin{axis}
        [scale only axis, width=\textwidth-3em-30pt, height=100pt,
         ymin = 0, enlarge x limits=false,
         ybar,
         xmajorgrids=false,
         scaled y ticks=false,
         xtick={},
         yticklabel={\pgfmathparse{\tick*100}\pgfmathprintnumber{\pgfmathresult}\%},
         bar width=1,
         ylabel={#2},
         xlabel={Resource ID},
        ]
\oddweekplot{#1}{#2}{1}
\evenweekplot{#1}{#2}{2}
\oddweekplot{#1}{#2}{3}
\evenweekplot{#1}{#2}{4}
\oddweekplot{#1}{#2}{5}
\evenweekplot{#1}{#2}{6}
\oddweekplot{#1}{#2}{7}
\evenweekplot{#1}{#2}{8}
\oddweekplot{#1}{#2}{9}
\evenweekplot{#1}{#2}{10}
\oddweekplot{#1}{#2}{11}
\evenweekplot{#1}{#2}{12}
\oddweekplot{#1}{#2}{13}
\evenweekplot{#1}{#2}{14}
\oddweekplot{#1}{#2}{15}
\evenweekplot{#1}{#2}{16}
\oddweekplot{#1}{#2}{17}
\evenweekplot{#1}{#2}{18}
\oddweekplot{#1}{#2}{19}
\evenweekplot{#1}{#2}{20}
\end{axis}
\end{tikzpicture}
}
\newcommand{\addtable}[2]{
\begin{figure}[b!]
{\addtikz{#1}{#2}}
\caption{`#2' for each resource in the #1 course}
\label{#1#2}
\end{figure}
}
\newcommand{\intaddtikz}[2]{
\begin{tikzpicture}
\begin{axis}
        [scale only axis, width=\textwidth-3em-30pt, height=100pt,
         ymin = 0, enlarge x limits=false,
         ybar,
         xmajorgrids=false,
         scaled y ticks=false,
         xtick={},
         bar width=1,
         ylabel={#2},
         xlabel={Resource ID}
        ]
\addplot
    table[x=ID, y=#2, col sep=comma]{#1Result.csv};
\end{axis}
\end{tikzpicture}
}
\newcommand{\intaddtable}[2]{
\begin{figure}[b!]
{\intaddtikz{#1}{#2}}
\caption{#2 learners for each resouce in the #1 course}
\label{#1#2}
\end{figure}
}

\begin{document}
\mainmatter

\author{Remi Brochenin\inst{1} \and Joos Buijs\inst{1} \and Mehrnoosh
Vahdat\inst{2} \and Wil van der Aalst\inst{1}}
\authorrunning{Brochenin et al.}

\institute{
Department of Mathematics and Computer Science,
Eindhoven University of Technology, 5612AZ Eindhoven, The Netherlands,
\and
Department of Industrial Design,
Eindhoven University of Technology, 5612AZ Eindhoven, The Netherlands\\
\{\email{r.brochenin, j.c.a.m.buijs, m.vahdat, w.m.p.v.d.aalst}\}\email{@tue.nl}
}

\title{Resource Usage Analysis from a Different Perspective on MOOC Dropout}
\titlerunning{Resource Usage Analysis}

\maketitle

\begin{abstract}
We present a novel learning analytics approach, for analyzing the usage of
resources in 
MOOCs. Our target stakeholders are the course designers who
aim to evaluate their learning materials.
In order to gain insight into the way educational resources are used,
we view dropout behaviour in an atypical manner: Instead of using it as an
indicator of failure, we use it as a mean to compute other features.
For this purpose, we developed a 
prototype, called RUAF, that can be applied to the data format provided by
FutureLearn.
As a proof of concept, we perform a study by applying this tool to the
interaction data of learners from four MOOCs. We also study the
quality of our computations, by comparing them to existing process mining approaches.
We present results that highlight patterns showing how learners use resources. We also show examples of
practical conclusions a course designer may benefit from.

\end{abstract}

\section{Introduction}
Learning Analytics (LA) and Educational Data Mining (EDM)
use data to inform and support the stakeholders about the learning
behaviour \cite{chatti2012reference,Vahdat2015esann2}. For instance, instructors can
gain insight into the performance of learners, and learners can
benefit from personalized guidance \cite{papamitsiou2014learning}.

Analysis of interaction data of learners with Massive Open Online Course (MOOC)
platforms is of growing interest for LA and EDM researchers.
Indeed, the automatic collection and availability of data has raised
interest in MOOCs \cite{baker2009state} for gaining a better insight into properties of the learning behaviour.
In this context, course designers are important stakeholders who are
responsible for designing and planning courses. Understanding how the learners
access and use resources, would help the course designers to adapt the learning materials to better fit the needs of learners.
Research on MOOC data often revolves around dropout behaviour and a
notion of success \cite{0.1}. This notion of success is based, for instance, on
the completion of a proportion of the tasks or on the grade obtained in quizzes and
assignments.
In this context, our work differs by relying on the fact that most of the
audience of MOOCs is already highly educated,
and many choose to study
only parts of the course resources \cite{yuan2013moocs}.
As a result, we do not direct our attention to the completion of the
course. Indeed, a learner who does not complete the course, but still spends
time viewing a variety of resources, will be considered as a valid learner.
Just like a textbook reader who would need the information of only a couple of
chapters, the MOOC learner can be selective about the course resources.
Considering this observation, we choose to focus on the
parts of the MOOC relevant for individual learners.

The main aim of this paper is to gain
insight into the resource usage behaviour of MOOC learners, by adopting this view on dropout.
For that purpose, we develop a prototype called RUAF (Resource Usage Analysis for FutureLearn)
to derive features about each resource, reflecting
how interesting this resource was for the whole audience, including those who
did not complete the course. For instance, we determine how many learners come
back to a resource for reference, or how many skip a resource. This prototype
can be applied to any of the MOOC datasets collected by the FutureLearn
platform. We made RUAF publicly available \cite{RUAF}.
Finally, we confirm the use of our prototype by testing our approach over four
MOOC datasets.
We also compare
our approach with the process
mining method of alignments.

This paper is structured as follows. In Section \ref{sectionRelated} we
present related work. Then, in Sections \ref{sectionData} and \ref{sectionMethod} we
present the datasets studied in this paper and explain
the RUAF prototype. We then demonstrate
the application of our prototype on the datasets in Section
\ref{sectionResults}. Finally, in Section \ref{sectionConclusions} we conclude
and indicate pointers for future work.

\section{Related Work}
\label{sectionRelated}
In this section we give an overview of related work in LA and EDM specifically targetting MOOCs.
After discussing the approaches most relevant to FutureLearn,
we take a more detailed look at questions of understanding learner behaviour and
then more precisely at questions about video usage.

\subsection{FutureLearn: A Growing MOOC Platform}
FutureLearn is a growing MOOC platform promoted by the Open University (UK).
Course designers are provided with access to the interaction data of their learners.
A variety of LA and EDM approaches have been applied to FutureLearn data \cite{0.1}. For
instance \cite{5.1}
provides some short insights into how FutureLearn, through developing
analytics dashboards, tries to provide a better feedback to stakeholders.

However, the reported works insist on looking at data in the same way as
traditional classrooms, in which the interactions between students are now
electronic and hence logged. The summary in
\cite{0.1} is confirming our diagnostic, with in general a traditional
view on dropout and the completion of a course, supporting a dichotomy
between success and failure.

\subsection{Analysis of Resource Usage in MOOCs}
MOOC usage behaviour from the perspective of the resources has not been studied
as often as the overall behaviour of individual learners. However, understanding
how the MOOC materials are accessed can be valuable and helps to improve
the quality of lessons and structure.

There has been some attempts at analyzing MOOC video interaction patterns to
identify the problems in the resources and their difficulty level for learners.
In \cite{1.2} a clickstream analysis is described that is based on the available types
of interactions with videos. This study focuses on perceived difficulty of videos
for the learners and video revisiting behaviours, and provides insight for the
course designers. For instance, they advise to reduce the information overload
in the lecture slides so that the less strong students can follow the
course better.

In \cite{1.1},  a visual analytical system is presented to help
educators gain insight into the learning behaviour through clickstream data from
Coursera. The study offers several types of visualizations that show the
difference of behaviour while viewing videos of the course.
Behaviours such as ``pause'' and ``play'' are measured every second of the
course. This visualization highlights the parts of each video that are viewed
more often, or where learners chose to pause.
This informs the course designers about
the parts of the videos learners are interested in.
Similarly, in 
\cite{1.4} the authors try to analyze video watching patterns through
the detection of in-video dropout and peaks of activity within a video.



\subsection{Learner Behaviour as a Process}

In the context of LA and EDM, the event logs of learners can be considered as a
temporal and ordered process. Some studies consider the
interaction data of learners as process data and analyse for example whether the learners follow a
planned curriculum.

For example, in \cite{trcka2009local}, methods of process mining are applied
and  a framework is introduced
to help educators analyze educational processes,
and facilitate real-time detection of curriculum violations.
Also, in \cite{3.1}, the authors quantify how well
the learners follow the order of the curriculum, from
the event log of the learner.
They compute for instance a
feature related to the frequency of events, as well as a delay between the
availability of a resource and the access to said resource.
Their purpose is to compare learner
behaviours.


In \cite{Mukalaetal} a more customized process mining approach
was presented. An innovative measurement of the way students watch
resources is introduced. A relation to
each resource is computed for each student: the student can have watched the resource either
on time, typically after the previous resource in the curriculum and before the
next one, or early, or late. The tool is based on alignments, which we will
describe later in this paper.






\section{MOOC Data}
\label{sectionData}
In this section, we describe the way data is provided by FutureLearn since the
aim of our prototype RUAF is to be used on data for any course using this MOOC platform. We
collected datasets through two courses offered by our university and we were
granted access to two more datasets provided by external institutions.

\subsection{Data Collection From FutureLearn}
On the MOOC platform FutureLearn, each course is offered with a weekly basis.
The weekly structure encourages the learners to follow a relatively linear
approach to the course, since the weekly resources are provided to the learners
once every seven days.
Each week contains a list of resources numbered as
`week number'.`resource number' (e.g. 1.2 is the second resource in the first
week).
Each resource can be one of the following types: a video, an article, a
discussion, a quiz, or an assessment-related item. The assessment-related items
are tests similar to a quiz, or peer-reviewed assignments.

Courses may be provided through several runs, for each of which a separate
dataset is provided by FutureLearn. The dataset contains varied data that
describe the interactions of the learners with the platform and other learners.
We focus here on the part of dataset that is called the `step activity', which
contains the temporal interaction data of learners with the MOOC resources and
resembles an event log.
Table \ref{table:t1} shows an extract of the event log by FutureLearn. For each
learner and each resource the learner has accessed, there is an entry
stating the first time the learner accessed that resource, and the last time
they completed that resource. 
Note that compared to other platforms (e.g.~Coursera), the data provided by the
FutureLearn platform is less detailed.

\begin{table}[t!]
\centering
\caption{An extract of event logs recorded by FutureLearn.
	}
	\begin{tabular}{|l|l|l|l|}
	\hline
	\textbf{~learner\_id~} & \textbf{~resource~} & \textbf{~first\_visited\_at~} & \textbf{~last\_completed\_at~} \\ \hline
~learner1 & 1.1 & ~2016-07-11 00:02:28 UTC~ & ~2016-07-11 00:12:54 UTC~ \\ \hline
~learner2 & 1.1 & ~2016-07-11 00:20:30 UTC~ & ~2016-07-11 00:22:55 UTC~ \\ \hline
~learner3 & 1.1 & ~2016-07-11 00:34:18 UTC~ & ~2016-07-11 00:35:46 UTC~ \\ \hline
~learner1 & 1.2 & ~2016-07-11 00:38:20 UTC~ & ~2016-07-11 00:40:24 UTC~ \\ \hline
	\end{tabular}
	\label{tbl:log}
	\label{table:t1}
\end{table}

\subsection{Datasets}
\subsubsection{ProM course}
We collected data of a FutureLearn course called ``Introduction to Process Mining
with ProM'', provided by
Eindhoven University of Technology (TU/e).\footnote{\url{http://www.futurelearn.com/courses/process-mining}}
This MOOC covers topics related to process mining and
focuses on the practical use of the ProM tool. The duration of the course is
four weeks and videos are the main resource type.

The first week is introductory, and the learners install and do basic analyses using ProM. 
The second and third weeks offer more advanced applications
of ProM. Finally, the last week is studying a dataset through an assignment
and discussions related to the assignment. For this week, no new topic is
introduced, and no video is included.


\subsubsection{Additional datasets}
We obtained FutureLearn data from two more courses provided by the University
of Twente.
We used these datasets to further test our approach. These datasets are as
follows.

\textbf{Nano course}: a FutureLearn course that is offered for a duration of
four weeks,
``Nanotechnology for Health: Innovative Designs for
Medical
Diagnosis''.\footnote{\url{http://www.futurelearn.com/courses/nanotechnology-health}}

\textbf{Ultra course}: a FutureLearn course that is offered for a duration of
six weeks,
``Ultrasound Imaging: What Is
Inside?''.\footnote{\url{http://www.futurelearn.com/courses/ultrasound-imaging}}

\subsubsection{Converted course} 
We obtained a larger fourth dataset based on the Coursera MOOC called ``Process Mining
-- Data Science in Action'' offered by TU/e. The duration of this course is six
weeks. We chose this dataset so as to test the scalability of RUAF.
We converted the detailed clickstream data provided by Coursera into the
less informative and coarser 
format of the FutureLearn `step activity' where only the
first and the last access to each resource are kept.

\section{RUAF: Resource Usage Analysis for FutureLearn}
\label{sectionMethod}
The aim of our work is to provide the MOOC designers with insight on the
usage of the resources of their courses considering all participants.
The novelty of our approach can be explained as follows.

Firstly, we note that \cite{2.3} suggests that learners may participate in the
course in their own preferred way while they may be classified as dropouts.
These participants have their own pace and selection of the materials, and might not
follow the structure of the course as planned by the course designer.
In other words, MOOCs are different
from a classroom where the students need to accomplish a certain percentage in
the assessment. Learners are considered mature enough to choose to not follow
the entire course, and only focus on sections of it.
Hence, we do not consider dropout learners as course failures, and do not
exclude them.



A second characteristic of our approach is the assumption, validated in the
next section, that learners tend to view resources from the beginning up to a
certain point. We call this point the `dropout point' that refers to the
resource after which the learner ceases following the course.
We do not consider `dropout point' as a negative term since knowing this point
under our assumption means knowing the part of the course that interests a
learner: from the beginning up until this point. We call our `dropout
point' assumption $DPA$.

This allows us to differentiate between two possible reasons a learner does not use a resource:
(1) the learner has not passed their `dropout point' and instead they are skipping the resource,
(2) the learner has passed their `dropout point' and is not active any more.
The second case is not reported as skipping a resource.

\begin{figure}[b!]
\centering
\includegraphics[trim=80 178 80 178, width=1\textwidth, clip]{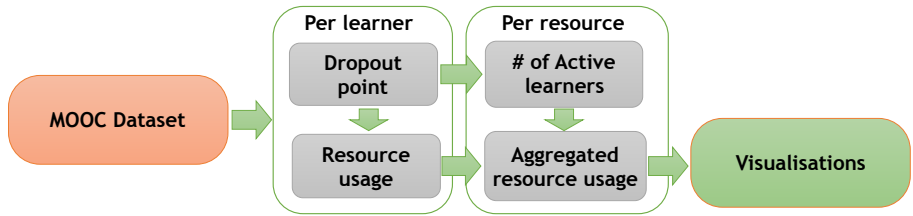}
\caption{RUAF prototype architecture.}
\label{imm:fig1}
\end{figure}

We develop a prototype, RUAF, which studies the behaviour of each learner
according to this view, and extracts
a set of features for each resource, as described in the following sections.
We made RUAF publicly available \cite{RUAF} for the research community.
The architecture of this tool follows the structure presented in Figure
\ref{imm:fig1}.

\subsection{Initial Preprocessing}
We consider a learner has \emph{done} a resource when
they interact more than a certain threshold in spending time with that
resource.
The threshold in our analysis is set to one minute. Thus, we exclude any learner
that has not spent more than one minute with any resource.


\subsection{Dropout Point}
The `dropout point' is the basis for the rest of our computations. We aim to
compute a reliable indicator of when each learner ceases to be interested in the
course.
For that purpose, we consider each learner independently from the others.
The `\emph{dropout point}' for each learner is defined as
the earliest resource such that: the learner has done
less than one third of the resources between that resource and any later resource.
This definition makes use of the assumption $DPA$ to attempt at capturing the
earliest point such that the learner is not involved in the remaining part of the course.


We define `dropout point' formally as follows.
Let $R$ be the total amount of resources in the studied course,
and $L$ be the set of learners. We define the function $D$ which for any learner $l\in L$
and any two integers $1 \leq i < r \leq R$ returns $D(l,i,r)$, the number of resources that $l$ has done
between the $(i+1)^{\textrm{th}}$ resource and the $r^{\textrm{th}}$ resource.
We also define the property $P(i,l)$ for an integer $i \leq R$ and a learner $l\in L$,
which holds if and only if for all $r$, if $i < r \leq R$ then $D(l,i,r)\leq (r-i)/3$. Finally,
the `dropout point' of a learner $l$ is
$\textrm{dropout}(l)=\min\{i,1\leq i \leq r \textrm{ and } P(i,l)\}$.

For instance, in a course with 9 resources, resource 3 is a good
candidate for the `dropout point' ($P(3,l)$ holds) if:
\begin{itemize}
  \item the learner has not done resources 4 and 5 (otherwise it would be more
  than one third of them).
  \item the learner has done at most one of the resources 4, 5, 6, 7 and 8.
  \item the learner has done at most two of the resources 4, 5, 6, 7, 8 and 9.
\end{itemize}

We can then aggregate this feature for all learners at the resource level.
We first compute for each resource how many learners are still active,
by counting how many learners have not passed their `dropout point' yet.
In other words, we consider that learners are active until their dropout point.
In mathematical terms, for the $r^{\textrm{th}}$ resource, the `\emph{active}'
feature is the number of $l\in L$ such that $\textrm{dropout}(l)\geq r$.

We compute the proportion of active learners who had their `dropout point' at a
particular resource, and obtain the feature we call `\emph{drop}'.
In more formal terms, for the $r^{\textrm{th}}$ resource, `drop' is the
number of $l\in L$ such that $\textrm{dropout}(l)=r$ divided by the
`active' feature of the $r^{\textrm{th}}$ resource.

The way we obtain the `dropout point' allows us to accurately exclude those
learners who are not interested in the course from a certain point while
keeping those who are interested but they are selective in using the resources.

\subsection{Usage Features}
From the `dropout point' we can determine for each learner and resource two
features. First, given a learner and a resource they overlooked, we
can know whether it was a skipped resource, or the learner dropped out before the resource.
We then aggregate at the resource level and divide that total by the number of
learners still active (the `active' feature of the resource), so as to obtain
the `\emph{skip}' feature: the proportion of active learners that have not done that
resource.
Then, similarly, given a learner and a resource they have done, we can know
whether the learner was simply active, or the learner peeked at a resource while
having already passed the `dropout point'. Then by aggregating at the resource
level and dividing by the total number of learners, we obtain the feature
about learners that `\emph{peek}' at a given resource (opposite of `skip').

We also compute features relevant to which order learners use to view resources.
First, we choose a threshold $k$ (two in our computations) for the minimum
number of resources that label the resource as done late or early.
We say that a resource $r$ is don late if at least $k$ resources were done
before $r$ while they should have been done after $r$ according to the
curriculum; and these $k$ resources are not good candidates at being done early
w.r.t. $r$. Formally, given a learner, we define two functions on
resource $r$ seen by this learner.
Let $A(r)$ be the set of resources that appear before $r$
in the curriculum, but that the learner has started after starting $r$.
Let $B(r)$ be the set of resources that appear after $r$
in the curriculum, but that the learner has started before starting $r$.
We say that:
\begin{itemize}
  \item $r$ was seen early: If there are at least $k$ resources $r_1,\ldots,r_k$
  such that for all $r_i$, both $r_i\in A(r)$ and the size of $B(r_i)$ is smaller
        than the size of $A(r)$;
  \item $r$ was seen late: If there are at least $k$ resources $r_1,\ldots,r_k$
  such that for all $r_i$, both $r_i\in B(r)$ and the size of $A(r_i)$ is smaller
        than the size of $B(r)$;
  \item $r$ was seen on time: Otherwise.
\end{itemize}

We then aggregate these notions at the resource level. We count for each
resource how many learners saw it late or early, divide the obtained figure by
the number of active learners, and obtain the features `\emph{late}' and
`\emph{early}'.

Finally, we compute whether a learner revisited a
resource, for instance for review or reference.
We count how many resources that appear later in the curriculum were
visited for the first time between the moment a particular resource was visited
for the first time and for the last time.
If the number of resources is more than a chosen threshold
(which we call the \emph{coming back threshold}, and set to three in our study),
we say that the learner came back to
this resource. Then we aggregate this at the resource level, counting for each
resource how many learners came back to it, and we divide this figure by the
number of active learners, and obtain the feature `\emph{back}'.

A summary of extracted features is presented in Table \ref{table:t2}.
\begin{table}[t!]
\centering
	\caption{Extracted features of a resource, and their description.
	}\begin{tabular}{|l|l|}
	\hline
	\textbf{feature~} & \textbf{description~}  \\ \hline
`active' & total number of learners who have not passed their `dropout point' yet\\
 \hline
`drop' & proportion of active learners in their `dropout point' at a particular\\
       & resource \\
 \hline
`skip' & proportion of active learners who skip a particular resource\\
\hline
`peek' & proportion of learners who have done a particular resource while they\\
  & have already passed their `dropout point'\\
\hline
`early' & proportion of active learners who have done a particular resource early\\
\hline
`late' & proportion of active learners who have done a particular resource late\\
\hline
`back' & proportion of active learners who come back to a particular resource\\
\hline
	\end{tabular}
	\label{tbl:features}
	\label{table:t2}
\end{table}

\subsection{Alignments}
Computing whether a resource is done late or early, as well as computing the
`dropout point', can be obtained through process mining techniques, with the
help of alignments. This method compares a process model with an event log, and
verifies if they match \cite{van2011process}.
In \cite{Mukalaetal}, a measure is introduced for determining if a resource is
done late or early with respect to an expected process model. We extended their
proposed method to be able to also compute the `dropout point'.

We create a process model (Petri net) representing the order of
resources set by the course designer
as in \cite{Mukalaetal}, with a modification. For each place situated after a
transition $r$ of the Petri net defined in \cite{Mukalaetal}, we add a
transition \mbox{drop-$r$} from that place to the end place. This
allows us to measure the dropout behaviour by computing alignments
to this
model.

We applied this modified tool to be able to compare our results to \cite{Mukalaetal}.

%
%
%

\subsection{Parameters}
RUAF can be tuned to the data of a particular course with these
parameters.

\textit{Minimum of time spent on a resource}: we consider that any learner
who spends less than one minute on a resource is the same as a learner who has
spent no time on that resource (and hence have not done that resource). This
time limit can be modified to any duration.


\textit{Dropout threshold}: the proportion of resources that a learner has not
done after the `dropout point'. We set this threshold to one third,
but this can be modified to any value. It can be useful to set a lower threshold if
the course contains multiple resources that are non-mandatory, such as
discussions, or links to further information.

\textit{Coming back threshold}: the threshold used to compute the `back'
feature, set to three in our analysis.

\textit{Early and late threshold}: the threshold
for being late or early ($k$ in the definition) is set to two in our
analysis.

\section{Results}\label{sectionResults}
In this section we present the results from applying RUAF to four
datasets, and compare our results to the method of
alignments in \cite{Mukalaetal}.\footnote{
Refer to Appendix \ref{Appendix} for all the visualizations provided by RUAF
applied to the four MOOC datasets, as well as a more detailed report of the
results from the method of alignments.}

\subsection{Application of RUAF to Four MOOC Datasets}
Here we present the results of the resource usage analysis with our prototype
RUAF for the four MOOC datasets introduced.

We only consider the learners who spent more than one minute on any
single resource.
With this view: the \textit{ProM course} had 908 learners, the \textit{Nano
course} had 935 learners, the \textit{Ultra course} had 2384 learners and the
\textit{Converted course} had 12026 learners.

\subsubsection{Drop}
We first study the `drop' feature, which can be seen as the dropout rate per resource.
According to literature, such as \cite{1.5}, we expected to see a much larger
dropout rate at the beginning of each course, decreasing throughout the course.
This expectation is confirmed in the ProM course, the Nano course and
the Converted course. However, the Ultra course disproves this hypothesis. The
`drop' feature is relatively stable throughout the course, with a few outliers.

In all the courses, an interesting pattern of `drop' emerges: at each transition from one week to the next, a
small peak of dropout occurs. It may be slightly before the end of the week or
slightly after the beginning of the next week, but is remarkably systematic.
This pattern is the most notable in the Ultra course, see Figure
\ref{UltraDrop} (note that as in all figures of the article, lighter
bars correspond to resources of even weeks).
The outliers of the Ultra course are all very close to a change of week, with
the exception of resources number 5 and 36.
\addtable{Ultra}{Drop}

An analysis of outliers in the dropout pattern shows a lack of interest in
assignments for a very large part of the learners. For instance,
a notable outlier is the last week of the ProM
course, during which the dropout rate suddenly increases. This week
exclusively contains an assignment and
discussions, resulting in only a third of the active learners to remain.


\subsubsection{Skip and Peek}
The feature `skip' can be seen as the proportion of learners still active at a
particular resource who chose not to pay attention to that resource.
The proportion of active learners who `skip' the video materials is quite
low.\footnote{In the case of the ProM and Converted courses, for which we
know the resource types.}  Figure \ref{ProMSkip}
shows the `skip' behaviour of the ProM course for all the materials. The
non-video materials (all long bars in Figure \ref{ProMSkip} are non-video), and
particularly articles, tend to be skipped much more than video resources. The
very high `skip' rate for articles can be explained as many are not mandatory.
For the videos, excluding three outliers, `skip' is
remarkably stable around 10\% for the first two weeks, then decreases by
half for the third week, while still being stable. The last week has no video.
As a comparison, in the
Converted course, `skip' is stable around 20\% for videos throughout the course.
\addtable{ProM}{Skip}



The `peek' behaviour shows a stable and very low rate for all resources, with
few exceptions. This observation is visible
in all four courses, for instance in the Nano course (see Figure
\ref{NanoPeek}) the `peek' varies from 0\% to 2\% for the majority of
resources. The outliers (all above 4\%) can be interesting for the designers of
this course. These resources may be an indicator of specific topics that the
learners who dropped out earlier had a particular interest in.
\addtable{Nano}{Peek}

Our analysis based on `skip' and `peek' features confirms the hypothesis that
learners access and use most resources until a `dropout point'. Indeed, the
learners have done on average about 90\% of videos up until their `dropout
point', and perform nearly no action after the computed `dropout point'. This
also indicates that our computation of `dropout point' is meaningful.


\subsubsection{Early, Late and Back}
The values for `early', as well as `late', are very low in all courses, except
the Ultra course.
The Ultra course, shown in Figure \ref{UltraEarly}, is characterized by higher
values of `early' (generally above 1\% from the second week) compared to the
other courses. Also, the high-valued outliers for the `early' feature are
generally at the beginning and end of the weeks. This may reflect the desire of
learners to know what remains in the course.
\addtable{Ultra}{Early}


In all courses, `back' is relatively stable, characterized by a decrease at
the end of each week. The ProM course, in Figure \ref{ProMBack}, is the only
one exhibiting a different behaviour. In its last week, the values of `back' sharply
increase. This can be explained by learners trying to find answers to their
final assignment by accessing to the discussions at the end of the week.
\addtable{ProM}{Back}

%
%

\subsection{Comparison with Alignments}
We compare our results with the work of \cite{Mukalaetal}, which uses
process mining to compute when a learner sees a resource early or late.
All the aggregated features we obtained from the alignments show the same
patterns as those from RUAF, which emphasizes that we are computing the same
notion.
Considering individual learners, we obtain results that are identical for the
large majority of the cases. There are some differences, most of which
are explained by two factors.

Firstly,
with the
alignment method from \cite{Mukalaetal}, as long as two items are
switched\footnote{Such as 3 and 4 in 1--2--4--3--5--6.}, one of them (chosen
arbitrarily) will be early or late.
On the other hand, we request more than two items not appearing in the expected
order so as not to need arbitrary choices.\footnote{For instance, in
1--2--4--3--5--6, it is unclear which of 3 or 4 is abnormal. Whereas in
1--4--2--3--5--6, resource 4 is seen early.}

Second, the `dropout point' is generally chosen earlier by alignments,
since the cost of handling unordered events may be higher than just considering that
the end of the process has been reached.\footnote{As an example, if the trace is
1--2--6--5--4, getting to resource 6 before dropout in an alignment would have
a higher cost  than choosing a dropout at resource 2.
}
This also leads to a difference in the computation of an aggregated `drop' from
the alignment results.
There are about 20\% less learners `active' per resource with this computation,
and on average learners have their `dropout point' more than five resources
earlier.

As a conclusion, although in the examples studied the obtained patterns in the
aggregated data are very similar, our approach has better results than
alignments for providing process-related insight into the data we study.
First, our computations never choose arbitrarily one possibility over another.
Second, having a complex non-linear behaviour does not encourage the algorithm
to set the `dropout point' earlier.

\section{Discussion and Conclusions}\label{sectionConclusions}
In the context of LA and EDM, there is a strong potential for progress in the feedback systems of MOOC
platforms \cite{0.1}. Developing ways to automatically analyze resource usage
and generate visualizations can be a valuable tool for course designers \cite{1.2,1.1}.
Course designers want their content to reach a large audience, 
this is not limited to people following the whole course.
From this perspective, learners who use just a part of the course
are as important as the certificate earners.
In this context, our work is a step towards considering all the users equally
when providing insight into the resource usage.

We developed a prototype, called RUAF, that measures the resource usage
properties through a novel approach toward `dropout point'. We made RUAF
publicly available \cite{RUAF} for the research community. We also studied the
quality of our computations, in particular by comparing them to what process
mining can compute and to the current statistics provided by FutureLearn
platform. We showed that our approach provides better results compared to
alignments computation of \cite{Mukalaetal}, while having much simpler semantics.

We finally showed in Section \ref{sectionResults} some examples of
practical conclusions a course designer may draw thanks to our prototype.
For instance, our prototype allows to detect which learning
resources are skipped or revisited by the learners, which resources provoke
dropout, which type of materials (articles, videos, assignment, etc.) are more
attractive for the learners, and which ones are accessed earlier or later than planned.

Such feedback is valuable for the course designers to tune the resources to the
needs of learners and direct their time and effort to the parts that need more
attention. For instance, if the order of resources in the curriculum is not
followed by learners, they can modify the order to have a more balanced
curriculum that is easier to follow. Also, detecting outliers in `skip',
`drop', and `back' can help them to exclude the resources that are problematic
for the learners, change their type, or reduce their difficulty level.

In the future, our approach can be implemented in the FutureLearn platform to
automatically recognize the resource usage properties, and provide feedback and recommendations to the course designers.
Additionally, we aim to extend our approach to analyze more properties of resource usage in particular in the case of MOOC
platforms that provide more detailed user
interaction data. For instance, we can
have a more detailed view on how students come back to a resource, how often they revisit resources and how they transit between the resources.
Finally, with very detailed data on video views (e.g. play and pause events), one could be able to pinpoint
particular patterns of usage within a video. For instance, within
a video there could be particular moments that provoke going back to another resource, or
skipping to the next resource. Course designers would then be able to relate
such information with the way a video is built, or with topic changes within a
video.

\bibliographystyle{splncs}
\bibliography{biblio.bib}

\clearpage
\appendix
\section{All visualisations}\label{Appendix}
We present here all visualisations of resource usage analysis
obtained through the work that led to this aticle. The subsections \ref{subsecappProM},
\ref{subsecappNano}, \ref{subsecappUltra}, and \ref{subsecappConverted}
are the visualisations of RUAF, exactly as the tool provides them. The
subsections \ref{subsecaliProM} and \ref{subsecaliConverted} present
features that are obtained through alignements.

\renewcommand{\addtable}[2]{
\begin{figure}[hb!]
{\addtikz{#1}{#2}}
\caption{`#2' for each resource in the #1 course}
\label{A#1#2}
\end{figure}
}

\renewcommand{\intaddtable}[2]{
\begin{figure}[hb!]
{\intaddtikz{#1}{#2}}
\caption{#2 learners for each resource in the #1 course}
\label{A#1#2}
\end{figure}
}

\newcommand{\wholecourse}[1]{
\subsection{Features for the #1 course}
\label{subsecapp#1}
Figure \ref{A#1Active} contains the graph of the number of learners that are
still active at each resource.

\intaddtable{#1}{Active}

Figures \ref{A#1Drop}, \ref{A#1Skip}, \ref{A#1Back},
\ref{A#1Early}, and \ref{A#1Late} contain respectively the graphs of the
features `drop', `skip', `back', `early', and `late' for each resource.
They are features computed as a
proportion of the number of active learners. Figure \ref{A#1Peek} contains the graph
of the feature `peek', which is computed as a proportion of all learners. In
these six figures, the change of week is shown by changing the shade of
blue. The even weeks are lighter than the odd weeks.

\addtable{#1}{Drop}

\addtable{#1}{Skip}

\addtable{#1}{Back}

\addtable{#1}{Early}

\addtable{#1}{Late}

\addtable{#1}{Peek}
}

\wholecourse{ProM}
\clearpage
\wholecourse{Nano}
\clearpage
\wholecourse{Ultra}
\clearpage
\wholecourse{Converted}
\clearpage

\newcommand{\addalitable}[2]{
\begin{figure}[hb!]
{\addtikz{#1-Alignment-}{#2-Alignment}}
\caption{Alignment-based `\lowercase{#2}' for each resource in the #1 course}
\label{ali#1#2}
\end{figure}
}

\newcommand{\alicourse}[1]{
\subsection{Features computed with alignments - #1 course}
\label{subsecali#1}
Figures \ref{ali#1Drop}, \ref{ali#1Early}, and \ref{ali#1Late}
respectively present features similar to the `drop', `early' and `late'
features, when obtained through the method of alignments. They are obtained from the same data as the features
presented in \ref{subsecapp#1}.

\addalitable{#1}{Drop}

\addalitable{#1}{Early}

\addalitable{#1}{Late}
}

\alicourse{ProM}
\clearpage
\alicourse{Converted}
\clearpage

\end{document}